\documentstyle[preprint,aps,epsf]{revtex} 
\input epsf 
\begin{document}

\draft 
\preprint{MSUPHY97.09} 
\title{Evaporation of a Kerr black hole
by emission \\ of scalar and higher spin particles}

\author{Brett E.\ Taylor\cite{BT}, Chris M.\ Chambers\cite{CC}, and
William A.\ Hiscock\cite{BH}}

\address{Department of Physics, Montana State University, Bozeman,
Montana 59717 }

\date{\today} 
\maketitle 
\begin{abstract} We study the evolution of an evaporating rotating
black hole, described by the Kerr metric, which is emitting either
solely massless scalar particles or a mixture of massless scalar and
nonzero spin particles.  Allowing the hole to radiate scalar particles
increases the mass loss rate and decreases the angular momentum loss
rate relative to a black hole which is radiating nonzero spin
particles.  The presence of scalar radiation can cause the evaporating
hole to asymptotically approach a state which is described by a
nonzero value of $a_* \equiv a / M$.  This is contrary to the
conventional view of black hole evaporation, wherein all black holes
spin down more rapidly than they lose mass.  A hole emitting solely
scalar radiation will approach a final asymptotic state described by
$a_* \simeq 0.555$.  A black hole that is emitting scalar particles
and a canonical set of nonzero spin particles (3 species of neutrinos, 
a single photon species, and a single graviton species) 
will asymptotically approach a nonzero value of $a_*$ only
if there are at least 32 massless scalar fields.  We also calculate
the lifetime of a primordial black hole that formed with a value of
the rotation parameter $a_{*}$, the minimum initial mass of a
primordial black hole that is seen today with a rotation parameter
$a_{*}$, and the entropy of a black hole that is emitting scalar or
higher spin particles.
\end{abstract} \pacs{}

\section{Introduction}
 
The evolution of evaporating black holes is a
process that has been studied in great detail.  Recently we found that
a black hole initially described by the Kerr metric which is radiating
only massless scalar particles via the Hawking process, 
asymptotically evolves towards a state
described by a rotation parameter $a \simeq 0.555 M$\cite{prl}.  
This is contrary to the conventional view of an
evaporating black hole's evolution which is that an initially rotating 
black hole will
spin down, evolving towards a Schwarzschild state before most of its
mass has been lost. For black holes emitting solely scalar radiation,
not only do initially rapidly rotating holes fail to spin down to
a Schwarzschild state, but holes with initial values $a < 0.555 M$
will actually spin up as they evaporate, again tending towards
an asymptotic state with $a \simeq 0.555 M$. Such a hole is losing  
angular momentum as it evolves, but is losing mass at a higher rate.

The evolution of evaporating black holes described by the
Schwarzschild and Kerr metrics was studied in detail by
Page\cite{Page1,Page2}.  In particular, Page numerically investigated
the evolution of a rotating uncharged hole described by the Kerr
metric that was emitting radiation via the Hawking
process\cite{Page2}.  He found that a black hole emitting nonzero spin
particles will lose angular momentum at a greater rate than it loses
mass.  Such a black hole will reach a state described by $a_* \equiv
a / M  \approx 0$ by the time it has lost approximately half its mass, 
allowing the
hole to be described by the simpler Schwarzschild metric in the late
stages of its evolution.  While Page only investigated radiation by
nonscalar fields in detail, he suggested that a hole that is emitting
massless scalar particles might evolve differently.  Page found that as the
black hole evaporated the dominant mode for the nonscalar
fields' radiation was that which had $l = m = s$, where $l$ and $m$ 
are the usual                                  
spherical harmonic indices and $s$ is the spin of the field.  Only a
scalar field can radiate in the $l = 0$ mode, and, by Page's argument,
one might expect this mode to be dominant for scalar radiation.  This
mode carries off energy from the hole, but no angular momentum.  If a
black hole is emitting scalar field radiation and the $l = 0$ mode for
that radiation dominates the overall emission from both scalar and
nonscalar fields, then the hole could lose mass faster than it loses
angular momentum.  It would then evolve asymptotically to a state
described by a nonzero value of $a_*$. This asymptotic value will 
hereafter be denoted by $a_{*0}$.  
This would mean that the hole
would always be described by the Kerr metric rather than evolving
asymptotically toward a Schwarzschild black hole.  We have shown that
this is the case if the black hole emits only scalar field
radiation\cite{prl}.

In this paper we discuss in more detail our results reported in
Ref.\cite{prl}.  We also determine whether or not the emission 
of massless scalar particles in addition to radiation from 
a canonical set of nonzero spin fields (3 separate neutrino, photon,
graviton) via Hawking radiation can
allow a black hole described by the Kerr metric to evolve towards 
an asymptotic state described by a nonzero value of $a_*$. 
We find that a single massless
scalar field in addition to the canonical set of nonzero spin 
fields makes very
little change in the evolution found by Page\cite{Page2} and that the
hole will still approach a state described by $a_* = 0$.  In order to
reach an asymptotic state described by  
$a_{*0}$ we find one must allow the
hole to emit radiation from a minimum of 32 massless scalar fields in
addition to the canonical nonzero spin fields.  

Our results combined with those of Page's results from Ref. \cite{Page2}
allow one to calculate the evolution of a Kerr black hole which is
evaporating with an arbitrary collection of scalar and nonzero spin
fields.
The results of our evolutionary calculations from both purely scalar 
radiation and scalar plus the canonical set of nonzero spin fields 
are applied to calculate a number of quantities of interest. These
include the lifetime of a Kerr black hole for a variety of initial
conditions, the minimum initial mass a 
primordial Kerr black hole could have possessed assuming it was 
observed today with a rotation parameter of $a_{*}$, the initial mass 
of a primordial Kerr black hole which would have just disappeared today, 
and the evolution of the area, or equivalently the entropy, of such 
holes as they evolve.  

In section II we discuss the mathematical formulae that
will be used to determine the evaporating black hole's evolution.  
In Section III the numerical methods that were
used are outlined.  The results of our work are presented in
Section IV, and a final summary offered in Section V.
Our notation follows that of Page\cite{Page2} and Teukolsky 
and Press\cite{TeuPress}. We use natural units, such that 
$G = c = \hbar = k_B = 1$, except where otherwise specified.

\section{Mathematical Formulae}

We will assume that the black hole is in isolation and that enough
time has passed so that the the black hole has lost all of its initial
charge (if any), but is rotating and will therefore be described
by the Kerr metric.  The wave equation for the massless scalar field,
$\Box \phi = 0$, separates in Kerr-ingoing coordinates 
\cite{TeuPress} by choosing
$\phi = R(r)S(\theta)e^{-i\omega \nu} e^{i m \tilde \phi}$, where the
angular function $S(\theta)$ is a spheroidal harmonic\cite{Flammer}.
The radial function $R(r)$ satisfies 
\begin{equation} 
	\left(\partial_r \Delta \partial_r - 2iK\partial_r - 2i\omega r 
	-\lambda \right)R(r) = 0 \; , 
\label{wave} 
\end{equation} 
where $\Delta = r^2 - 2Mr + a^2$, $K =
(r^2 + a^2)\omega - am$, $\lambda = E_{lm\omega} - 2am\omega +
a^2\omega^2$ and $E_{lm\omega}$ is the separation constant.  A general
solution to Eq.\ (\ref{wave}), expressed in terms of known functions,
is not known, but asymptotic solutions can be found\cite{TeuPress},
\begin{eqnarray} 
R \longrightarrow \left\{ \begin{array}{ll} Z_{{\rm
	hole}} & r \rightarrow r_{+} \\ Z_{{\rm in}} r^{-1} + 
	Z_{{\rm out}}r^{-1}e^{2 i \omega r} & r \rightarrow \infty 
	\end{array} \right.  \; .  
\label{asymptote} 
\end{eqnarray} 
The subscript ``in'' refers to an
ingoing wave originating from past null infinity, ``out'' refers to an
outgoing wave reflected from the hole that propagates toward future
null infinity, and ``hole'' refers to the component of the wave that
is transmitted into the black hole through the horizon at $r = r_+$.
The amplification, $Z$ (the fractional gain of energy in a scattered
wave), is 
\begin{equation} 
	Z = \left\vert {Z_{\rm out} \over Z_{\rm
	in}} \right\vert^2 - 1.  
\label{amplification} 
\end{equation}

Following Page\cite{Page1} we express the rates at which the mass and
angular momentum decrease by the quantities $f \equiv
-M^2 dM/dt$ and $g \equiv -M a^{-1}_* dJ/dt$ respectively, which have been
scaled to remove overall dependence on the size (mass) of the black
hole. The coordinate $t$ is the usual Boyer-Lindquist time coordinate.
These quantities will
determine the evolution of the black hole.  They are defined by
\begin{equation} 
	\left( \begin{array}{c} f \\ g \end{array} \right)
	= -\sum_{l,m} \frac{1}{2\pi} \int^{\infty}_{0} dx 
	\frac{Z}{e^{2 \pi k / \kappa} -1} 
	\left( \begin{array}{c} x \\ m a_{*}^{-1} \end{array}
\right) \; , 
\label{fandg} 
\end{equation} 
where $k = \omega -
m\Omega$, $\Omega = a_*/2r_+$ is the surface angular frequency,
$\kappa = \sqrt{\left(1 - a_*^2\right)}/2r_+$ is the surface gravity
of the hole, and following Page\cite{Page2} we have defined $x =
M\omega$.  The relative magnitude of the mass and angular momentum loss
rates will determine whether or not the hole will spin down to a nonzero
value of $a_*$.  To describe how the angular momentum and mass loss
rates compare we define 
\begin{equation} 
	h(a_*) \equiv {d \ln a_*\over d \ln M} = 
	{g(a_*) \over f(a_*)} - 2.  
\label{h} 
\end{equation}

To investigate how the mass and angular momentum evolve in time, it is
convenient to define new quantities.  Again following
Page\cite{Page2}, we define 
\begin{equation} 
	y \equiv - \ln a_* \; ,
\label{y} 
\end{equation} 
which we will use as a new independent
variable replacing $t$.  If the initial mass of the black hole
at $t = 0$ is defined to be $M_1$, then a dimensionless mass
variable $z$ may be defined by 
\begin{equation} 
	z \equiv -\ln(M/M_1) \; ,
\label{z} 
\end{equation} 
which has the initial value $z(t = 0) = 0$.
The hole's mass, now parameterized by $z$, with our definitions of $f$
and $g$ as well as Eq.\ (\ref{y}) and Eq.\ (\ref{z}), will then evolve
according to 
\begin{equation} 
	{dz \over dy} = {1 \over h} = {f \over
	g-2f} \; .  
\label{zvsy} 
\end{equation} 
Finally we define a
scale-invariant time parameter 
\begin{equation} 
	\tau \equiv M_1^{-3} t \; , 
\end{equation} 
with the initial value $\tau(t = 0) = 0$.  The
evolution of $\tau$ with respect to $y$ is then determined by
\begin{equation} 
	{d\tau \over dy } = {e^{-3z} \over g - 2f} \; .
\label{tauvsy} 
\end{equation} 
To see explicitly how $a_*$ evolves with
respect to time $t$ we can use Eq.  (\ref{tauvsy}) to find
\begin{equation} 
	{da_* \over dt} = - {a_* h f \over M^3}\; .
\label{dadt} 
\end{equation} 
At points in the evolution where $g = 2f$, or equivalently where $h = 0$,
Eq.\ (\ref{tauvsy}) is numerically bad.  This is due to the fact that
the hole will be described by a constant value of $a_*$ as it continues
to lose mass.
For Eq.\ (\ref{dadt}) we see that since the black hole is in isolation, it
can only lose mass (via the Hawking process), so the function $f$ must
be positive definite.  Therefore, if there is a nonzero value of $a_* = 
a_{*0}$, for which $h$ is zero, then $da_* / dt $ will be zero 
there and the hole will remain at that value of $a_*$. If $d h/ d a_*$ 
is positive at such a point, then it represents a stable state towards
which holes will asymptotically evolve. If $d h/ d a_*$
is negative or zero at a point where $h = 0$, then it  
represents an unstable equilibrium point of $a_*$, and holes will evolve
away from it. We will show that emission of scalar radiation may 
create stable states of $a_*$ at points where $h=0$ and $d h /d a_*
> 0$, depending on the mix of fields present. 

The evolution of the mass and angular momentum of the hole are 
completely determined by Eq.(\ref{zvsy}) and Eq.(\ref{tauvsy}).
For collections of fields for which the black hole spins down 
completely, evolving towards a state described by a value of 
$a_* = 0$, the situation resembles that analyzed by Page. 
First, the initial point of the numerical integration  will be taken to 
be nearly extremal with $a_* = 0.999$. The resulting functions
$z(y)$ and $\tau(y)$ may then be used to describe evolution from
other initial rotations $a_{*i}$, with the initial values
\begin{equation}
	y_i \equiv - \ln(a_{*i})  ,
\label{yidef}
\end{equation}
\begin{equation}
	z_i \equiv z(y_i) = - \ln(M_i/M_1)  ,
\label{zidef}
\end{equation}
\begin{equation}
	\tau_i \equiv \tau(y_i) = {M_i}^{-3}t_i  .
\label{tauidef}
\end{equation}
A hole which began with $a_* = 0.999$ and $M=M_1$ at $t=0$ will
then have $a_* = a_{*i}$ and $M=M_i$ at $t=t_i$. 

For a collection of fields for which the hole evolves to a spinning
asymptotic state described by $h(a_{*0}) =0$
(as happens, for example, with purely scalar radiation) two 
distinct evolutionary cases must be investigated.  One of these is 
to start out again with a nearly extremal hole as described above, 
where now the possible initial values $a_{*i}$ range only from the 
initial near extremal value to the asymptotic limiting value $a_{*0}$
for that set of fields.  In the second case the initial point of the 
evolution will be taken to be nearly Schwarzschild with an initial 
value $a_* = 0.001$.  In this case the value of $a_*$ will increase 
as the evolution progresses, increasing towards the asymptotic value 
$a_{*0}$. The resulting evolution may be used to describe all initial 
states with $a_{*i}$ ranging from the initial value up to the 
limiting value $a_{*0}$ for that set of 
fields.

Once we have determined the evolution of the hole there are a number
of other quantities we would like to investigate.  The first is how
the area, and hence the entropy, of the hole evolves.  This can be
done by calculating the area at each step of the evolution, 
\begin{equation} 
	A = 8 \pi M^2 \left[1 +\left(1 - a_{*}^2\right)^{1/2}
	\right]\; .  
\label{area} 
\end{equation}

Another quantity of interest is the lifetime of a black hole 
that is described by an initial value of $a_* = a_{*i}$ and just 
evaporates today.  A scale invariant lifetime $\theta$ can then 
be given by 
\begin{equation} 
	\theta _i = e^{3z_i}\left(\tau_f - \tau_i \right)\; . 
\label{lifetime}
\end{equation} 
The quantity $\tau_f$ is the scale invariant time required to reach the 
endpoint of the evolution from the maximal initial values of
the numerical integration.
In Ref.\cite{Page2}, Page defined $\tau_f$ by $\tau_f \equiv 
\tau(y = \infty)$.  Note that as $y \rightarrow \infty$, $a_* 
\rightarrow 0$.  Since a black hole emitting scalar particles
may not spin down to $a_* = 0$, we instead use a more general
definition, namely that $\tau_f \equiv \tau(z = \infty)$.  
In other words, we define the lifetime in terms of the time it
takes the mass to reach zero, rather than the time until $a_* = 0$ 
(since $a_*$ may not ever approach zero in the general
case).  Here $\tau_i$ is the time it takes for a hole to evolve
from an initial value of $a_* = a_{*i}$ and mass $M_i$ to $M = 0$.
Once the lifetime of a hole is known, one can then calculate the 
initial mass of a primordial black hole formed with $a_* = a_{*i}$ 
that has just evaporated within the present age of the universe $t_0$:
\begin{equation} 
	M_i(a_{*i},t_0) = t_0^{1/3} \theta_i^{-1/3} =
	t_0^{1/3}e^{-z_i}\left(\tau_f - \tau_i \right)^{-1/3}\; .
\label{massprimordial} 
\end{equation}

If a primordial black hole were observed today with rotation parameter 
$a_{*}$, a minimum value on its initial mass could be found by
letting $\tau_i$ go to its minimum value.  For a black hole that evolves 
by spinning down, either to a nonrotating state, or towards an asymptotic
state described by $a_{*0}$, the minimum value is simply 
given by
setting $\tau_i = 0$, provided primordial black holes can be formed 
with initial values of $a_* = 1$.  The other case is when 
the hole forms with an initial value of $a_*
< a_{*0}$, so that it spins up as it evolves.  In this
case the limiting value of $\tau_i = 0$ again, but we must choose a 
limiting value for the initial value of $a_*$.  Here we choose the 
initial value to be $a_{*} = 0.001$, which is larger than our numerical
error.  The minimum mass for either the spinup or spindown case is 
then given by
\begin{equation} 
	M_{\rm min}(a_{*},t_0) = t_0^{1/3}e^{-z}\tau ^{-1/3}\; .
\label{massminimum} 
\end{equation}

\section{Numerical Methods}

The bulk of the numerical calculation is computing the amplification
for each mode according to Eq.\ (\ref{amplification}). The asymptotic 
forms of the radial part of the wave equation for the massless scalar 
field given in Eq.\ (\ref{asymptote}) can be used to numerically integrate
Eq.\ (\ref{wave}) and obtain the function $R$.  We followed Bardeen's
method of integration described in Teukolsky and Press\cite{TeuPress}.
Initial conditions were chosen to be a purely ingoing wave at the event
horizon using the asymptotic form of $R$ as $r \rightarrow r_+$ found
in Eq.\ (\ref{asymptote}).  This solution was then integrated outward
to a large value of $r$ where it was resolved into its ingoing and
outgoing components.  Note from Eq.\ (\ref{asymptote}) that the ingoing
and outgoing solutions both fall off as $1/r$ as $r \rightarrow
\infty$. This creates a difficulty as both modes will contribute
to the same order to the numerically integrated value of $R$.  To
separate the contributions from each mode note that a derivative of
$R$ as $r \rightarrow \infty$ gives 
\begin{equation} 
	{dR \over dr} \simeq -{R \over r}+{2i\omega Z_{\rm out} 
	e^{2i\omega r} \over r}\; .
\label{dRdr} 
\end{equation}
Since $dR/dr$ was already calculated in our numerical integration we can 
use this value and the value of $R$ to calculate $Z_{\rm out}$ and then 
$Z_{\rm in}$.  The amplification due to scattering can then be 
calculated from Eq.(\ref{amplification}).  This must be done for each 
value of $l$, $m$, and $\omega$.  This numerical integration was done 
using a Bulirsch-Stoer\cite{NR} method to an accuracy of one part
in $10^4$ for the entire integration.  
We found that the accuracy of the value calculated for the
amplification depended on the endpoint of the radial integration.  If
the integration did not proceed to a large enough value of $r$ some of
the scattering would be missed.  For this reason a check was made to
insure that the integration had proceeded to a large enough value of
$r$ such that the value for the amplification was constant to within a
part in $10^4$ as well.

A final numerical integration was done to determine how the mass and
angular momentum of the hole evolve in time.  This was done by
integrating Eq.\ (\ref{zvsy}) and Eq.\ (\ref{tauvsy}).  Since $f$ and
$g$ are necessary for the integration, both were calculated at 12
values of $a_*$ so as to match up with the data given by Page for the 
nonzero spin fields.  We fit the values for $f$ and $g$ for the scalar
field by a 10th order polynomial in $a_*$.  This fitting resulted 
in a standard deviation for both $f$ and $g$ on the order of a part in
$10^4$.  A similar fit could be obtained by using a 4th order polynomial, 
however the standard deviation was larger.  Fitting of the nonzero spin
fields was done in the manner prescribed by Page\cite{Page2}.
The fitting for $f$ and $g$ is done to save computation time by 
eliminating an additional
integration for $f$ and $g$ as the integration for Eq.\ (\ref{zvsy}) 
and Eq.\ (\ref{tauvsy}) proceeded.  The integration for $f$ and $g$ 
has an overall accuracy of approximately a part in 100.

Total values for $f$ and $g$ may be found by summing the contributions 
over all of the fields 
\begin{equation}
	\left( \begin{array}{c} f \\ g \end{array} 
	\right) = 
	\sum_{s} {\rm n}_s 
	\left( \begin{array}{c} 
		f_{s} \\ g_{s}
		\end{array} 
	\right)\; ,
\label{fandgtotal} 
\end{equation} 
where $s$ indicates the spin of the field and ${\rm n}_s$ is the number 
of fields with that spin.  One can find the mass and angular momentum 
loss rates for an arbitrary collection of fields by using 
Eq.\ (\ref{fandgtotal}) together with the scalar field data provided 
in Table \ref{table1} and the results for the higher spin fields 
found in Table I of Ref.\ \cite{Page2}.  

The final integration of Eq.\ (\ref{zvsy}) and Eq.\ (\ref{tauvsy}) was
done using an adaptive stepsize, fourth order, Runge-Kutta routine to an
accuracy of a part in $10^4$ per step.  The total number of steps for a 
typical integration was a few hundred, resulting in an overall accuracy 
of approximately a part in 100.  The values of $z$, or equivalently
$M/M_{\rm 1}$, and $\tau$ were then used to calculate the area,
lifetime, the initial mass of a primordial black hole that has just
evaporated, and the minimum initial mass of a primordial black hole
seen today with a value of $a_{*i}$ over the range $0 \leq a_{*i} \leq
1$.  

Numerically there is a difficulty with Eq.\ (\ref{lifetime}) for small
initial masses $M_i$.  Note that 
as ${\rm M}_{\rm i} \rightarrow 0$ the quantity $z_i \rightarrow \infty$.
At the same time, the quantity $\tau_f - \tau_i$ is going to zero.  
It is difficult to accurately determine the evolution numerically
here, and it is useful to instead use an analytic approximation.  Here
Page used a power series expansion to determine the lifetime when $a_*
\rightarrow 0$, due to his definition of $\tau_f$.  Since in our case
$a_*$ may not tend to zero, we had to follow a different procedure.
Using L'H\^opital's rule one can obtain the limiting behavior when
$M \rightarrow 0$ and find 
that it agrees with Page's approximation of the lifetime when $a_*$ 
is small\cite{Page2}.  
\begin{equation}
    \theta_i = \lim_{M_i \rightarrow 0} {1 \over 3 f(a_{*i})}\; .
\label{limitinglife}
\end{equation}

\section{Results}

\begin{figure} \epsfxsize = 276pt \epsfbox{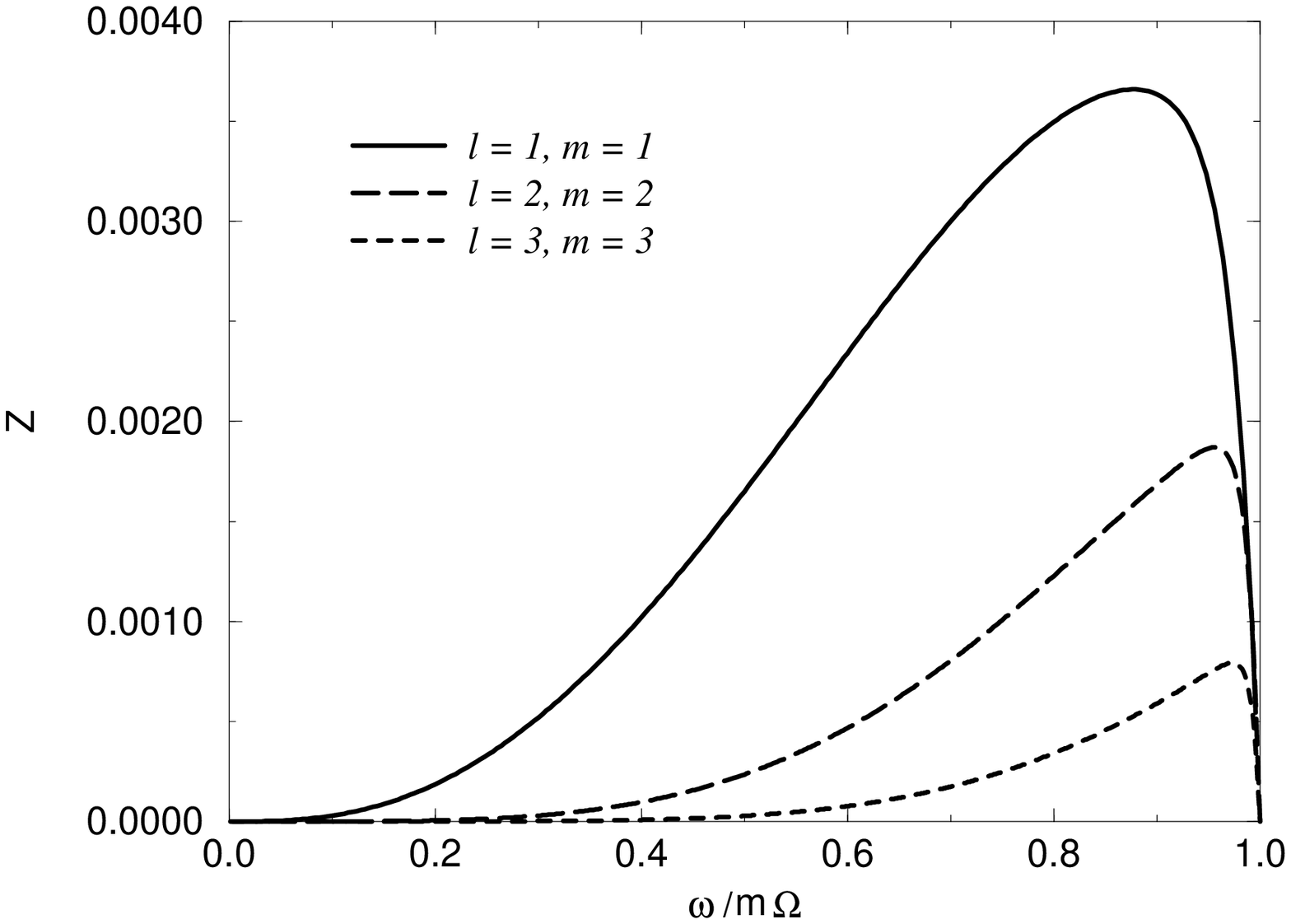} 
\caption{The
amplification Z of massless scalar radiation from a Kerr black hole
with rotation parameter $a_* = 0.99$, plotted in the superradiant
regime, $0 \leq \omega \leq m\Omega$.  Note the $l = 0$ mode is not
superradiant and is therefore not shown.}  
\label{amp} 
\end{figure}

In 1972 Press and Teukolsky calculated the amplification of scalar
waves from a Kerr black hole\cite{PTNature}.  They noted that
there was a problem with their numerical code as it was generating
nonzero values for Z at $\omega = m\Omega$, which is the upper limit of
the superradiant regime.  In subsequent papers\cite{TeuPress,PTold}
they calculated the amplification for spin 1 and spin 2 fields.  In
both cases they note that the maximum amplification occurs for
those modes which have $l = m = s$.  Spin 1/2 fields exhibit no
superradiance due to the fact that they are fermions and obey the
Pauli exclusion principle.  For all nonscalar fields, the $l = 0$ mode
is a non-radiant mode.  For a scalar field, this mode is radiant but
does not exhibit superradiance.  Sample amplification curves in
the superradiant regime for a massless scalar field are shown in
Fig.\ \ref{amp}.  It can be seen that all of the curves go to $Z
= 0$ at $\omega = m\Omega$ as they should.  To the best of our knowledge
this is the first time the amplification for a scalar field scattering
off a Kerr black hole has been calculated correctly.  To accomplish this we
used the Bardeen transformation found in\cite{TeuPress} 
which Press and Teukolsky used for the 
spin 1 and spin 2 fields. 
The amplification can 
be used to calculate $f$ and $g$, the scale invariant mass and angular
momentum loss rates respectively.

\begin{figure} 
\epsfxsize = 276pt \epsfbox{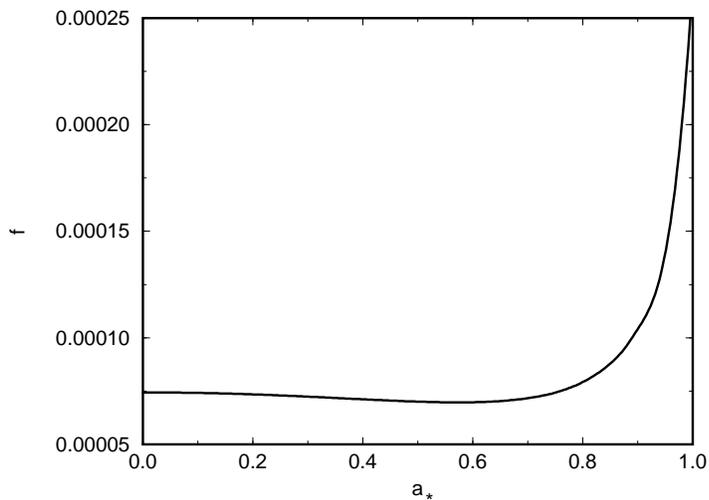} 
\caption{The scale
invariant mass loss rate is shown versus the rotation parameter for a
single massless scalar field.  There is a minimum located at $a_* =
0.574$.  In contrast, for nonzero spin fields the function $f$ is monotonically
increasing.}  
\label{f} 
\end{figure}

Page found that for the nonzero spin fields $f$ is a monotonically
increasing function of $a_*$\cite{Page2}.  However for the scalar
field this is no longer true.  As we noted in \cite{prl} and as shown
in Fig.\ \ref{f}, for a scalar field $f$ has a minimum located
approximately at $a_* \simeq 0.574$.  The existence of this minimum 
is due to the
fact that, unlike the nonzero spin fields, the scalar field can
radiate in an $l = 0$ mode and this mode dominates $f$ at low values
of $a_*$.  The emission in this mode then decreases as $a_*$ increases.  
This suggests that the $l = 0$ mode couples more strongly to the hole 
at low values of $a_*$.  Emission in the superradiant modes however 
monotonically increase as $a_*$ increases.  These two effects combine 
to form the observed minimum.  For both scalar and nonzero spin fields 
$f$ is a positive definite function, indicating that the hole's mass
always decreases into the future.  This is expected since we assumed 
that the hole was in isolation so there would be no way for it to 
gain mass.  Using a tenth order polynomial fit we can extrapolate our 
numerical values for $f$ to a $a_* = 0$ and compare with known results 
for Schwarzschild black holes.  We find that $f \rightarrow 7.44 
\times 10^{-5}$ as $a_* \rightarrow 0$, which agrees to three 
significant digits with the previous results found by Simkins
\cite{Simkins} and Elster\cite{Elster}.

\begin{figure} 
\epsfxsize = 276pt \epsfbox{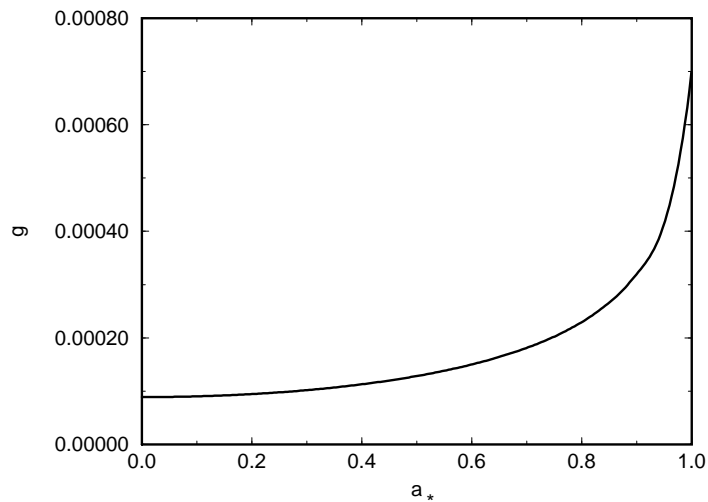} 
\caption{The scale
invariant angular momentum loss rate is shown versus the rotation
parameter $a_*$ for a single massless scalar field.  The curve is
qualitatively similar to those for the nonzero spin fields, being a
monotonically increasing function of $a_*$.}  
\label{g} 
\end{figure}

For the scale invariant angular momentum loss rate, Page found for
nonzero spin fields that $g$ is a monotonically increasing function of
$a_*$\cite{Page2}.  As shown in Fig.\ \ref{g} this is also true for the
scalar field.  The reason the scalar field result is qualitatively
similar to the nonzero spin fields is that the $l = 0$ mode carries
off no angular momentum so it makes no contribution to $g$.  The form
of $g$ can then be understood purely in terms of superradiance, which
causes $g$ to increase as $a_*$ increases.  A table of values for
$f$ and $g$ for a massless scalar field can be found in Table
\ref{table1} which complement those given for the nonscalar fields by
Page\cite{Page2}.

\begin{figure} 
\epsfxsize = 276pt \epsfbox{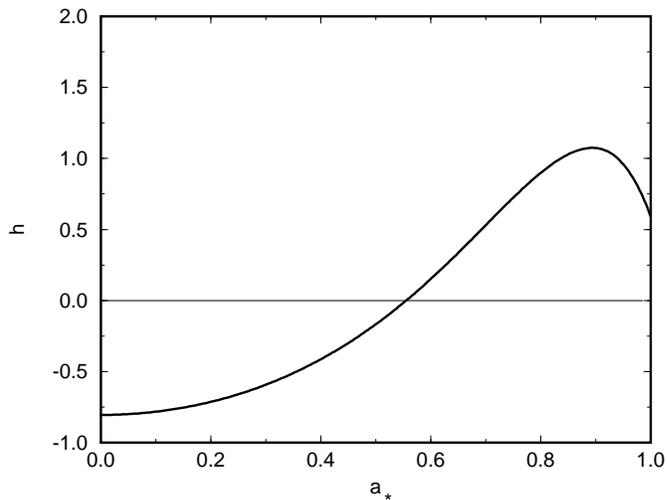} 
\caption{The scale
invariant quantity $h$ which describes the rate of change of the
angular momentum relative to that for the mass is plotted versus the
rotation parameter for a single massless scalar field.  The function
$h(a_*)$ has a zero at $a_* \simeq 0.555$.  A hole that forms with 
$a_* > 0.555$ will then spin down to that value as it evolves
while one that forms with $a_* < 0.555$ will spin up towards 
that value.} 
\label{hfig}
\end{figure}

Using our results for $f$ and $g$ we can calculate $h$ from Eq.\
(\ref{h}).  Note from Eq.\ (\ref{dadt}) that $da_*/dt$ will be zero only
when $h = 0$, since $f$ is positive for all $a_*$ for a hole
in isolation.  Fig.\ \ref{hfig} shows the behavior of $h$ due
solely to particle emission by a scalar field.  The most important
feature is that $h(a_*) = 0$ at a value of $a_* \simeq 0.555$ as seen
in the figure and noted in\cite{prl}.  A black hole that forms with a
value of $a_* > 0.555$ will have $h(a_*) > 0$ so by Eq.\ (\ref{dadt}),
$da_*/dt$ will be negative and the value of $a_*$ will decrease
as the hole evaporates, tending towards $a_* = 0.555$.  In contrast 
a hole that forms with $a_* < 0.555$ will have $h(a_*) < 0$ so 
$da_*/dt$ will be positive and the value of $a_*$ will increase 
towards $a_* = 0.555$.  Thus, when only emission of scalar field 
particles is considered, a black hole with any nonzero initial 
angular momentum will evolve towards a state with $a_* \simeq 0.555$.  
This should be compared to what was found for nonzero spin fields 
where the hole rapidly evolves to a state characterized by $a_* = 0$.

Page found that the value of $h$ at $a_* = 0$ satisfied an 
approximate linear relationship with the spin for the nonzero spin 
fields he examined:
\begin{equation} 
	h_s(a_* = 0) \simeq 13.4464s - 1.1948 \;.  
\label{hvss}
\end{equation} 
Extrapolating this to $s = 0$ indicates that $h_0(a_* = 0)$ is negative, 
which led Page to suggest that emission of scalar particles might 
prevent a black hole from spinning down.  While this conclusion is 
confirmed by our calculations, we find $h_0(a_* = 0)= -0.806$, which 
does not satisfy Eq.(\ref{hvss}), indicating that the approximate 
linear relation breaks down for the scalar case.

\begin{figure} 
\epsfxsize = 276pt \epsfbox{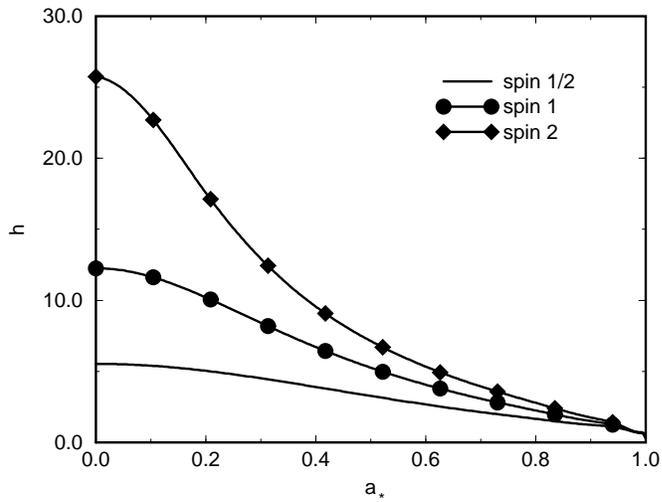} 
\caption{The
scale invariant quantity $h$ is plotted against the rotation parameter
for each of the nonzero spin fields.  The function for each of the
fields is positive definite for all values of $a_*$, showing that a
Kerr black hole emitting radiation from nonscalar fields loses 
angular momentum faster than mass.}  
\label{hmix} 
\end{figure}

As seen in Fig.\ \ref{hmix}, $h$ is a positive definite function for
the nonzero spin fields.  Since we have found for a scalar field that
$h$ has a region in which it takes negative values it seems clear
that whether $h=0$ at some nonzero value of $a_*$ depends on the 
collection of fields present in nature or considered in a model
problem.

We can now numerically evaluate Eq.\ (\ref{zvsy}) and Eq.\ (\ref{tauvsy}) and
determine how a rotating hole's mass and angular momentum will evolve
with solely a scalar field and with a mixture of fields.  For simplicity,
we will ignore emission by massive fields. We will consider cases
including massless neutrinos, and also cases without neutrino 
contributions. We choose sets of species that correspond to sets 
chosen by Page\cite{Page2} to facilitate comparison.   
Within the standard model the only fundamental scalars are the Higgs 
bosons, which have a relatively high mass.  By the time the temperature
is high enough to emit these massive scalar particles, many other higher 
spin particles will be produced in substantial numbers and the overall
situation will become overly complicated due to possible symmetry
restoration.  For this reason the scalar fields will be taken to be 
truly massless in our discussions.

\begin{figure} 
\epsfxsize = 276pt \epsfbox{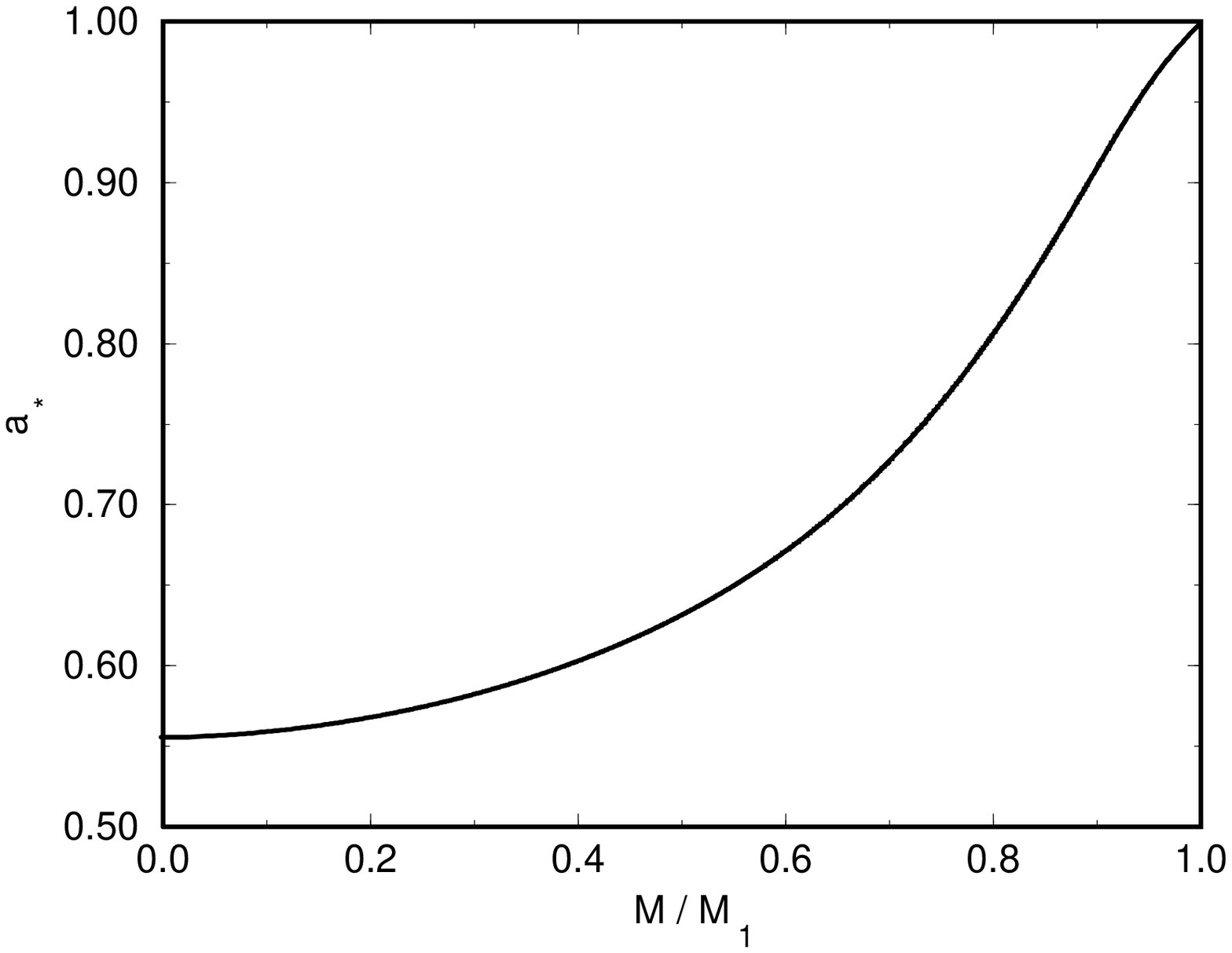}
\caption{The mass of a black hole evaporating solely by emission of 
radiation from a
single massless scalar field is shown plotted versus its rotation
parameter $a_*$ for an initially rapidly rotating hole.  
The black hole evolves to a state characterized by
$a_* \simeq 0.555$ from an initial state characterized by $a_* = 0.999$ and
initial mass $M_1$.  The evolution of a black hole that has a
different initial value of $a_* = a_{*i}$, but in the range $0.555 <
a_{*i} \leq 1$ can be found by locating the desired value of $a_{*i}$ 
on the curve and rescaling the horizontal axis so that
$M/M_1 = 1$ at that point.}  
\label{mvsascalar1} 
\end{figure}

The first case we will consider will be that of a black hole emitting
radiation from a single scalar field as it evaporates.  In such a
case, the hole will evolve towards a state specified by $a_{*0}
\simeq 0.555$.
As seen in Fig.\ \ref{mvsascalar1}, a hole that starts out nearly
extreme with a value of $a_* = 0.999$ has reached a state
characterized by $a_{*0}$ by the time it has lost 85 \% of its
initial mass.  As $a_* \rightarrow a_{*0}$, we find that $a_* - 
a_{*0}$ is
proportional to the mass of the hole squared.

\begin{figure} 
\epsfxsize = 276pt \epsfbox{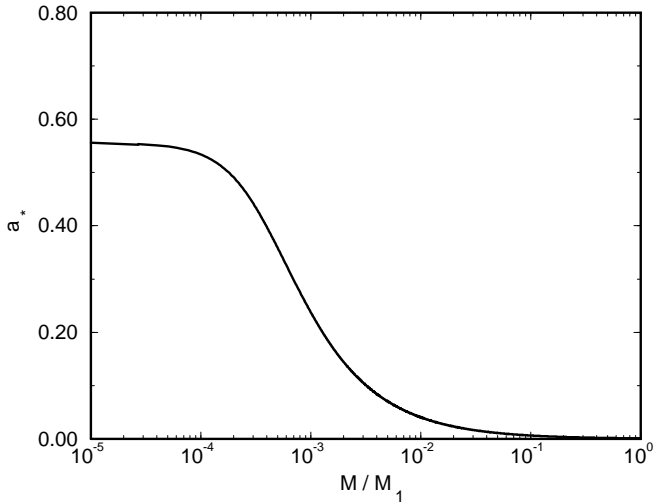}
\caption{The mass of a black hole evaporating solely by emission 
of radiation from a single massless scalar field is shown plotted 
versus its rotation
parameter $a_*$ for an initially slowly rotating hole.  
The black hole evolves to a state characterized by
$a_* = 0.555$ from its initial state characterized by $a_* = 0.001$.
The evolution of a black hole that has a different initial value of
$a_* = a_{*i}$, but in the range $0 \leq a_{*i} < 0.555$ can be found
by locating the desired value of $a_{*i}$ on the curve
and rescaling the horizontal axis so that $M/M_1 = 1$ at that point.}
\label{mvsascalar2} 
\end{figure}

In contrast a hole that starts out nearly nonrotating, with 
$a_* = 0.001$, stays at roughly the same value of $a_*$ until
it has only approximately 1 \% of its initial mass. At that point its
value of $a_*$ increases and reaches $a_{*0} \simeq 0.555$ once it has
about 0.01 \% of its initial mass remaining.  This evolution
is shown in Fig.\ \ref{mvsascalar2}.  As $a_* \rightarrow a_{*0}$,
we again find that $a_*-a_{*0}$ depends quadratically on the mass.

The second case we consider is one in which the hole is emitting
particles from a single massless scalar field and, in addition, from the
known massless higher spin fields.  Specifically we will allow the
hole to emit particles from a single spin 1 and spin 2 field in
addition to a single scalar field.  This changes the evolution only 
slightly from that found by Page; in particular, the hole still 
appears to spin down completely.

The third case we describe involves massless spin 1 and spin 2 fields 
and in addition three spin 1/2 fields representing the neutrino flavors.  
This set will be called the canonical set of fields.  In addition to the
canonical set we also allow a single massless scalar field to be
present.  The resulting evolution is very similar to that found by 
Page for the canonical set of fields; the hole again rapidly spins
down to a Schwarzschild state. The three additional neutrino fields 
have a very small effect on the evolution, as these fields are not 
superradiant.

Since the addition of a single scalar field is seen to have little
effect when the multiple known fields of nature are included in the
mix, we next determined how many scalar fields would be necessary in
addition to the canonical set of fields to allow the hole to evolve
towards a state described by nonzero $a_{*}$. 
Given the canonical set of fields, 
we find it takes a minimum of 32 massless scalar fields to cause 
the hole to evolve to a state with $a_{*}$ nonzero.  This combination
of fields will evolve to a state described by $a_* = 0.087$.
This indicates that if there were at least 32 massless scalar fields 
in nature, then evaporating black holes could evolve towards an 
asymptotic state described by a nonzero value of $a_{*}$.  It is unlikely
this will be realized, since there are at present no known 
massless scalar fields in nature.

\begin{figure} 
\epsfxsize = 276pt \epsfbox{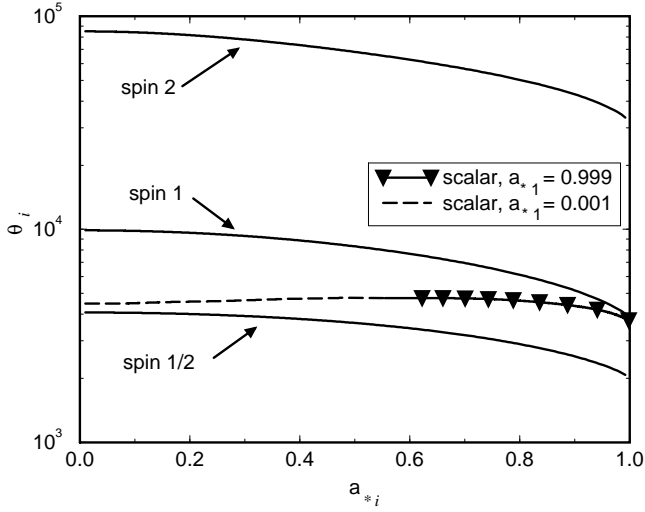}
\caption{The scale invariant lifetimes for primordial black holes
emitting radiation from differing single massless fields is plotted
versus the initial value of the rotation parameter, $a_{*i}$, that the black
hole formed with.  Here we see the two distinct scalar evolutions, one 
which spins up from a nearly Schwarzschild state and one which spins down 
from a nearly
extreme state, both asymptotically approach $a_* \simeq 0.555$.  
In this and the following figures, the two different evolutionary
paths due
to emission of purely scalar particles from the black hole are differentiated 
by the use of a symbol on one of the curves.  The number of symbols is not
related to the number of data points and their use is intended only to help 
clarify the figure.}  
\label{lifescalar} 
\end{figure}

In Fig.\ \ref{lifescalar} we see the scale invariant lifetime given by
Eq.  (\ref{lifetime}) for a primordial black hole that was formed with
rotation parameter $a_{*i}$, which has just disappeared and which evaporated
by purely massless radiation into a single species.  

\begin{figure} 
\epsfxsize = 276pt \epsfbox{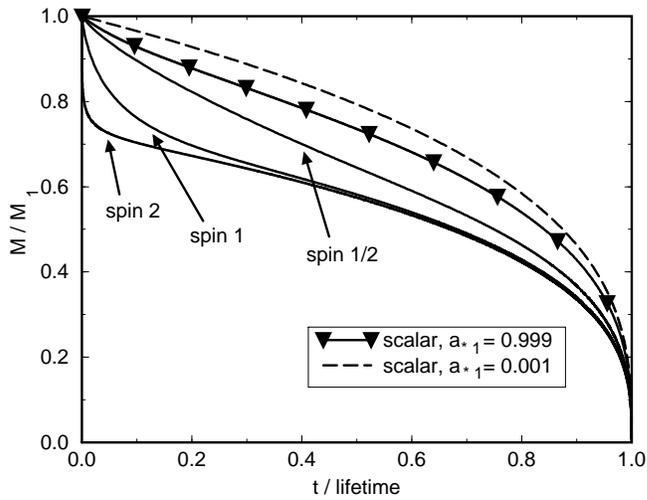}
\caption{The fractional mass is plotted against the fractional
lifetime for a black hole that is emitting radiation from a single
species.  Pure massless scalar radiation decreases
the mass loss rate relative to the nonzero spin fields.  The two different
scalar curves represent a hole that is starting in a nearly extremal state
or a nearly Schwarzschild state.}
\label{mvst_species} 
\end{figure}

In Fig.\ \ref{mvst_species} the fractional mass is plotted versus the
fractional time for holes that are emitting purely in one species.
The emission due to a single scalar field allows the hole to lose mass
more slowly than in those cases which emit nonzero spin particles.  Here 
the nonzero spin fields each carry off more energy than a single scalar field.

\begin{figure} 
\epsfxsize = 276pt \epsfbox{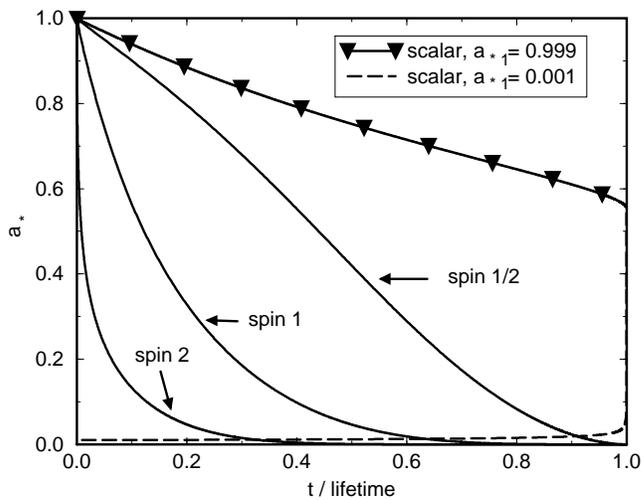}
\caption{The rotation parameter is plotted versus the fractional time
for a black hole that is evaporating by radiation from a single field.
Massless scalar radiation causes the hole to spin down more
slowly than in the nonzero spin cases.  The evolution of a black hole
that starts out with a different value of $a_{*} = a_{*i} $ can be
found by shrinking the vertical axis from the top for those holes 
starting at a nearly extremal state to the desired value
of $a_{*i}$ and rescaling the horizontal axis so that $t_{\rm initial}
= 0$.  The same can be done for the hole which begins nearly 
Schwarzschild and is radiating into a single massless scalar field 
by the above process, but starting from the lower left corner.}
\label{avst_species} 
\end{figure}

In Fig.\ \ref{avst_species} the rotation parameter is plotted versus
the fractional lifetime for holes that are emitting purely in a single
species.  As before we see that purely scalar emission causes the hole
to approach an asymptotic value of $a_* \simeq 0.555$, either from
above or below, depending on the initial state of the black hole.  

\begin{figure} 
\epsfxsize = 276pt \epsfbox{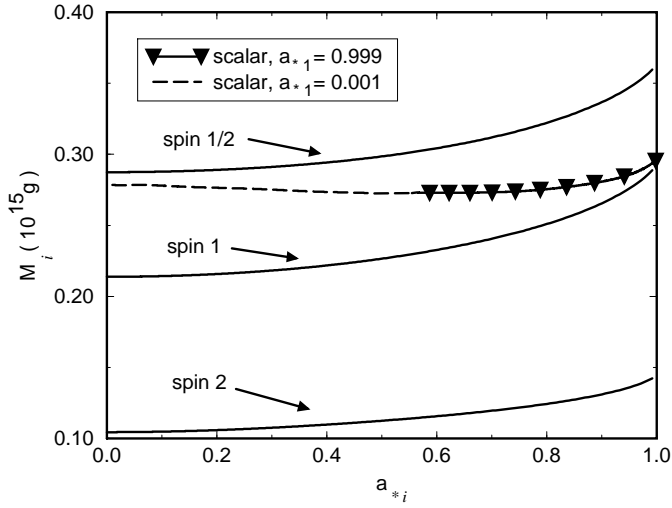}
\caption{The initial mass of a primordial black hole that has just
disappeared within the present age of the universe by emission of
radiation from a single massless field is plotted versus the value of the
rotation parameter when it formed, $a_{*i}$.  Here the two distinct evolutions
by emission of purely scalar particles, one spinning up and one spinning down,
can be seen to converge on a state described by 
$a_* \simeq 0.555$.}  
\label{prim_sc}
\end{figure}

In Fig.\ \ref{prim_sc} the initial mass of a black hole that has just
disappeared today via radiation from a single massless field is
plotted against the initial value of its rotation parameter $a_{*i}$.
As Page did, we assume the age of the universe is $t_0 = 16
\times 10^9$ years which implies in our units that $t_0^{1/3} = 4.59 \times
10^{15}$ g.  
We used the limiting form of the lifetime shown in Eq.\ (\ref{limitinglife})
to calculate the lifetime as $M_{\rm i} \rightarrow 0$.

\begin{figure} 
\epsfxsize = 276pt \epsfbox{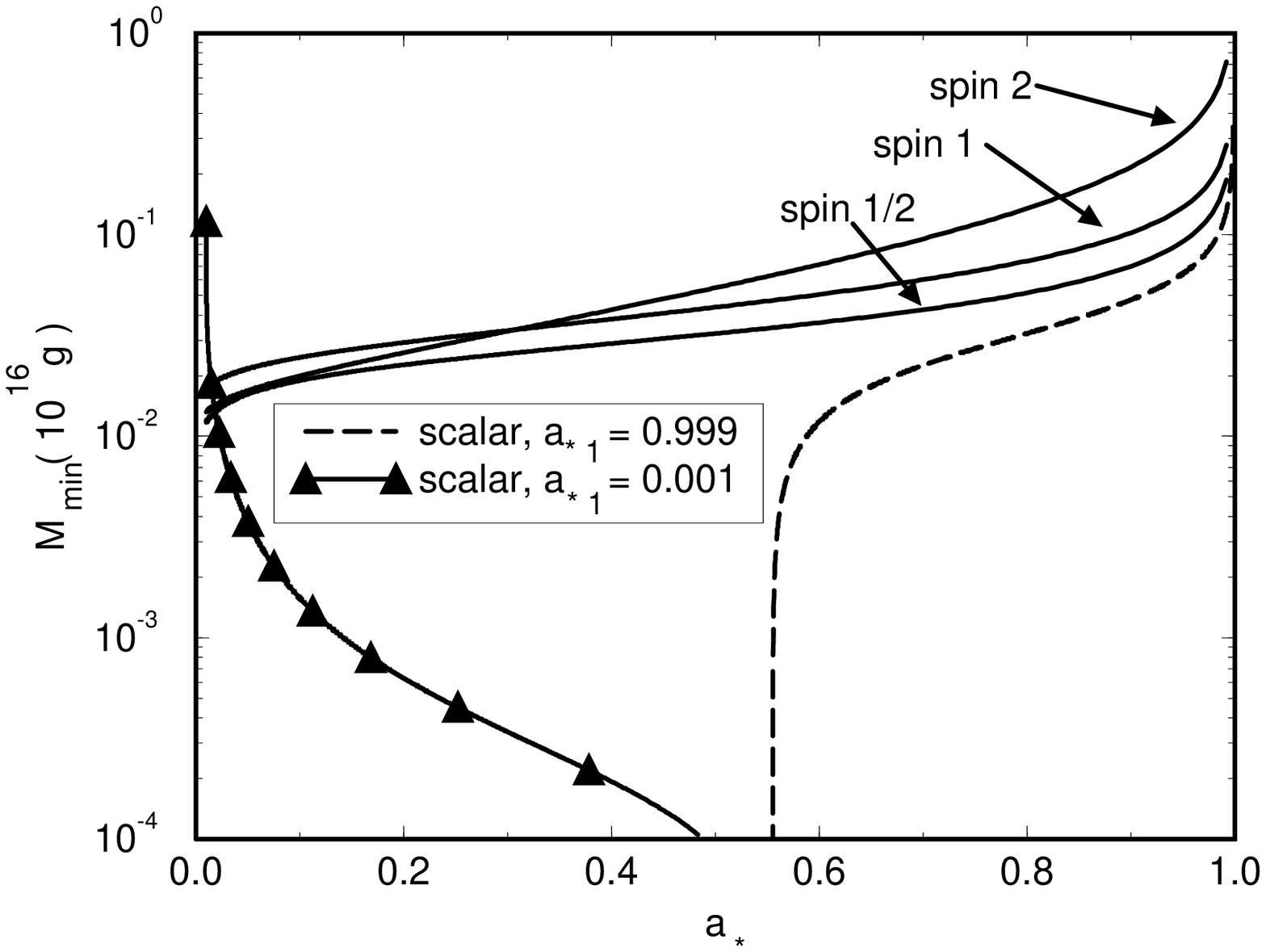} 
\caption{The
minimum mass of a primordial black hole that has been seen today is
plotted against the value of the rotation parameter that it has today,
$a_{*i}$ for black holes emitting radiation from single massless fields
.  For a hole that is emitting purely scalar particles, both the initially
near extremal and near Schwarzshild evolution curves asymptotically approach
a state characterized by $a_* \simeq 0.555$.}  
\label{min_mass} 
\end{figure}

In Fig.\ \ref{min_mass} the minimum mass that a primordial hole could
have had at the time of its formation is plotted against the value of 
the rotation parameter $a_{*}$ it has today.  Here the two scalar
curves representing the differing spinup and spindown evolutions
show significantly different behavior, from each other and from
the higher spin fields. In particular, note that if only massless
scalar emission occurred, a hole observed with $a_*$ slightly greater
than 0.555 would have a significant uncertainty in its initial mass,
due to the steepness of the spin 0 curve shown here.

\begin{figure} 
\epsfxsize = 276pt \epsfbox{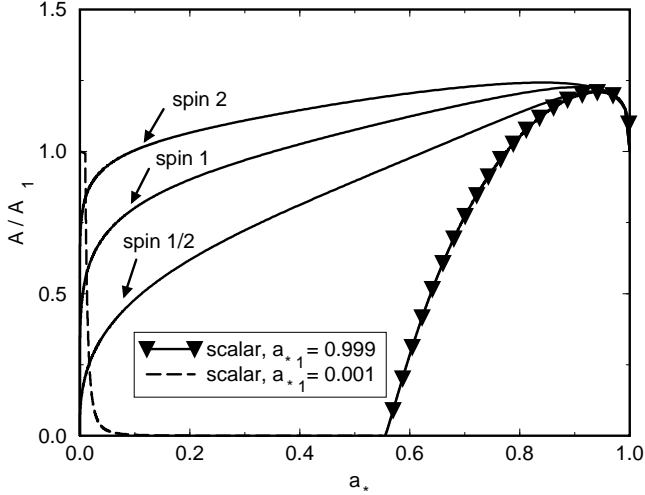} 
\caption{The
fractional area is plotted versus the rotation parameter for a black
hole which is emitting radiation from a single field.  Pure scalar
radiation causes the area to decrease more rapidly (as a function of
$a_*$) than for the nonzero spin fields, particularly for initially slowly
rotating holes.  The evolution of the area from a
black hole that starts out with a value of $a_{*i} > 0.555$  can
found by shrinking along the horizontal axis from the right 
hand curve to insure that $A /A_{\rm 1} = 1$.  For a hole that is 
emitting only scalar particles and which 
forms with $a_{*i} < 0.555$ the same process can be followed, only 
shrinking from the left curve rather than the right.
This set of curves also represents the evolution of
the entropy.}  
\label{area_sp} 
\end{figure}

In Fig.\ \ref{area_sp} the ratio of the horizon area to the
initial area is
plotted versus $a_{*}$ for each of the individual fields. Essentially,
$a_{*}$ is being used here as an alternate time parameterization to
spread out the rapid evolution which occurs near $a_{*} =1$. 
Note that the area actually increases initially as $a_*$ decreases
for nearly extreme holes.  The evolution of the area can be
described by differentiating Eq.\ (\ref{area}) 
\begin{equation} 
	{dA\over dt} = {A \over M^3} \left({hf \over (1 - 
	a_*^2)^{1/2}} -g\right)\; .  
\label{area_evol} 
\end{equation} 
Since $h$ and $f$ are both positive as $a_*\rightarrow 1$ the first
term on the right hand side of Eq.(\ref{area_evol}) dominates, causing 
the initial increase in
the area.  Since the entropy is one quarter the horizon area these 
results also describe the evolution of the black hole's entropy.

\section{Discussion}

In this paper we have described how the emission of massless scalar
field radiation via the Hawking process affects the evolution of a Kerr
black
hole in isolation.  We find that for low values of $a_*$ the hole's
dominant emission mode for the scalar field is the $l = 0$ mode.  
This follows the same trend as the nonzero spin fields, namely that 
the $l = m = s$ mode dominates the emission.  
We found that the scale invariant mass loss
rate $f$ has a minimum located at a value of $a_* \simeq 0.574$ for the
massless scalar field while in contrast Page found that for the nonzero 
spin fields $f$ is monotonically increasing with $a_*$.  This is due 
to the fact
that the $l = 0$ mode is not a radiant mode for the nonzero spin
fields and this mode dominates the emitted scalar radiation when
$a_*$ is small.  As $a_*$ increases the higher $l$ modes, with $m \neq 0$, 
which are
superradiant, begin to increase in strength while the $l=0$ mode emission
decreases due to poor coupling with the hole at high values of $a_*$.
Together these two effects combine to form the minimum.  We also found
that the scale invariant angular momentum loss rate for the scalar
field is qualitatively similar to the results Page found for the nonzero spin
fields, being a monotonically increasing function of $a_*$.  Combining
these two results we showed that the quantity $h$ which describes the
rate at which the angular momentum and mass of the hole are changing
relative to each other reaches a value of zero at $a_* \simeq 0.555$ for a
hole emitting purely massless scalar radiation.  This implies that, for the
evolution of a black hole-quantized scalar field system, a
black hole which possesses any nonzero initial angular momentum will evolve 
towards an asymptotic Kerr black hole state described by this value of $a_*$,
which we argue is a stable point of the evolution.  By comparing our value 
of $h(a_* = 0)$ with that predicted by Page's hypothesized linear 
relationship for the nonzero spin fields we have shown that the proposed
relationship breaks down for the scalar case.

Using our results for $f$ and $g$ for the scalar  
field along with the values of $f$ and $g$ given by Page for the nonzero 
spin fields, one can calculate the evolution of a Kerr black hole evolving
by the emission of a collection of particles of arbitrary spin.
We have calculated the evolution of a Kerr black hole for a number of
cases in which the hole is allowed to emit radiation from different fields.  
For a hole that is emitting only radiation from a single massless scalar
field in addition to radiation from the known massless fields 
(electromagnetic and gravitational), the evolution follows much the same
trend as that found by Page for purely nonzero spin fields.  We also
find that the neutrino fields have very little effect on the evolution
due to the fact that they are not superradiant.  If we allow
the hole to emit radiation from the two known massless nonzero spin fields
and three neutrino fields we find that the hole will asymptotically
approach a nonzero value of $a_{*}$ only if there are 32 or more massless
scalar fields present.  

We also found that the scale invariant lifetime is decreased when a
hole is allowed to radiate massless scalar particles in addition to
the canonical set due to the addition of new radiative channel.  
The initial mass of a black hole that
has just now evaporated correspondingly is thus increased, although only
slightly. We also found that by allowing the hole to emit into
a massless scalar field in addition to the canonical set of fields
slowed the rate at which angular momentum is lost by the hole.  The
minimum mass for a hole that is seen today with rotation parameter
$a_{*i}$ deviates in a very insignificant matter whether or not the
hole is allowed to radiate into the canonical or canonical plus scalar
set of fields.  Finally the area, and consequently the entropy, was
found to be lower at each value of $a_{*i}$ for emission into the
canonical plus scalar set of fields versus just the canonical set of
fields, again due to the addition of another radiative channel in the 
form of the scalar field.

\acknowledgements

The work of WAH and BT was supported in part by NSF Grant No.
PHY-9511794.  CMC is a fellow of The Royal Commission For The
Exhibition Of 1851 and gratefully acknowledges their financial
support.  The authors would also like to acknowledge helpful
discussions with Don Page.

\newpage

\newpage

\newpage

\newpage

\newpage

\begin{table} 
\caption{The quantities of $f$ and $g$ for a single
massless scalar field are shown for 12 values of $a_*$.  The values of
$a_*$ were chosen to match those used by Page for the nonzero spin
fields.}  
\begin{tabular}{ccc} 
$a_*$&$f$&$g$\\ 
\tableline
0.01000&$7.429\times 10^{-5}$&$8.867\times 10^{-5}$\\
0.10000&$7.442\times 10^{-5}$&$9.085\times 10^{-5}$\\
0.20000&$7.319\times 10^{-5}$&$9.391\times 10^{-5}$\\
0.30000&$7.265\times 10^{-5}$&$1.024\times 10^{-4}$\\
0.40000&$7.097\times 10^{-5}$&$1.125\times 10^{-4}$\\
0.50000&$6.996\times 10^{-5}$&$1.281\times 10^{-4}$\\
0.60000&$7.008\times 10^{-5}$&$1.507\times 10^{-4}$\\
0.70000&$7.119\times 10^{-5}$&$1.803\times 10^{-4}$\\
0.80000&$7.969\times 10^{-5}$&$2.306\times 10^{-4}$\\
0.90000&$1.024\times 10^{-4}$&$3.166\times 10^{-4}$\\
0.96000&$1.551\times 10^{-4}$&$4.515\times 10^{-4}$\\
0.99000&$2.283\times 10^{-4}$&$6.160\times 10^{-4}$\\
0.99900&$2.625\times 10^{-4}$&$6.905\times 10^{-4}$\\
0.99999&$2.667\times 10^{-4}$&$6.997\times 10^{-4}$\\ \end{tabular}
\label{table1} \end{table}

\end{document}